\documentclass[prl,twocolumn,floatfix,showpacs,superscriptaddress]{revtex4}
\usepackage{graphicx}
\usepackage{floatflt}
\usepackage{amsmath}
\begin{document}

\title{Anomalous Kondo Spin Splitting in Quantum Dots}
\author{J. M. Aguiar Hualde}
\affiliation{Departamento de F\'{\i}sica, Facultad de Ciencias
Exactas, Universidad de Buenos Aires, Ciudad Universitaria, Pabellon
I, Buenos Aires, 1428, Argentina.}
\author{G. Chiappe}
\affiliation{Departamento de F\'{\i}sica Aplicada, Universidad de
Alicante, San Vicente del Raspeig, Alicante 03690, Spain.}
\affiliation{Departamento de F\'{\i}sica, Facultad de Ciencias
Exactas, Universidad de Buenos Aires, Ciudad Universitaria, Pabellon
I, Buenos Aires, 1428, Argentina.}
\author{E.V. Anda}
\affiliation{Departamento de F\'{\i}sica, Pontificia Universidade
Cat\'olica do R\'{\i}o de Janeiro, 22452-970 R\'{\i}o de Janeiro,
Caixa Postal 38071, Brasil.}

\date{\today}
\begin{abstract}

The Zeeman splitting of localized electrons in a
quantum dot in the Kondo regime is studied using a new
slave-boson formulation. Our results show that the Kondo peak
splitting depends on the gate potential applied to the quantum dot and on the topology of the
system. A common fact of any geometry is that the differential susceptibility shows a strong non
linear behavior. It was shown that there exist a critical field above which the Kondo resonance is splitted out.
This critical field rapidly diminishes when the gate potential is lowered, as a consequence of the reduction
of the Kondo temperature and a subsequent strong enhancement in the differential susceptibility occurring at low fields.
The critical field is also strong depedent on the topology of the circuit.
Above this critical field the Zeeman splitting depends linearly upon the magnetic field and does not extrapolate
to zero at zero field. The magnitude of the Y-intercept
coordinate depends on the gate potential but the slope of this function is not renormalized being
independent of the value of the gate potential. Our results are in agreement with very recent experiments.

\end{abstract}
 \maketitle The observation of Kondo effect
in quantum dots has already a large history \cite{gogo98,crocro98}. 
The effect appears when a net
spin becomes localized at a quantum dot (QD) coupled with metallic
leads \cite{glara98,ngle88}. Local spin fluctuations (LSF) in the
vicinity of the QD allows the spin of the electrons in the leads to
screen the localized spin, as a magnetic impurity is screened in a
host metal. A sharp resonance in the local density of states (LDOS)
is developed at the Fermi level of the system. When the QD is
connected as a bridge between two metallic leads, the embedded
geometry (EG), the resonance provides a channel of conduction and
the conductance reaches the value $G= 2G_0$, being $
G_0 = \frac{e^2}{h}$ the quantum of conductance. When the QD is side
connected (SCG), two channels of conduction interfere destructively
(one direct and other across the QD, from lead to lead) and the
conductance is zero for the value of the gate potential, for which
the interference is complete. The existence and characteristics of
the Kondo resonance can be studied through conductance $G$ of the system.

To better understand the Kondo effect it is
important to study its behavior when an external magnetic field is
applied. It is expected the LSF to be quenched and the Kondo regime
to disappear for large enough fields. For intermediate values, the
Kondo resonance is splitted and its effect can be observed in the
differential conductance of the system \cite{crocro98}. In 
recent experiments \cite{koam04,amge05} it was possible to obtain
very precise value of $\Delta$, the Zeeman splitting, (ZS) for
localized electrons in a QD in the Kondo regime (KR). However, these
experiments are controversial. In one of them, \cite{crocro98},
above a critical value of the magnetic field, it was found that
$\Delta \sim \Delta_B=2 g \mu_BB$, where $g$, $\mu_B$ and $B$ is the
gyromagnetic factor, the Bohr magneton and the magnetic field
respectively. In others, \cite{koam04,amge05}, bellow a critical
value, the effect of the magnetic field was not observed and above
it the splitting was larger than $\Delta_B$ and does not extrapolate
to zero at zero field.

The splitting was initially theoretically predicted \cite{meme93} to
be $\pm g\mu_BB$, in line with
\cite{crocro98}. However, more recent calculations \cite{co00} shows
that the splitting would be observable only above $B_c$, when the
magnetic Zeeman energy $\Delta_B$ becomes competitive with the Kondo
temperature $T_K$. Moore and Wen \cite {mowen00} established the
relationship between the field induced spin splitting in the
spectral function of the system in equilibrium and the splitting
appearing in the differential conductance when the system is
slightly out of equilibrium. They predict also that the spin
screening characteristic of the KR should reduce the magnetic field
splitting, ${\it i.e}$ $\Delta<\Delta_B$. Unfortunately these
theories were not capable of explaining the phenomenology seen in
the latest experiments \cite{koam04,amge05}.

In this letter we developed a theoretical study that correctly
explains these last experimental results. We use
different approaches to study the problem. We propose a new
slave-boson formalism(SB) that considers two independent boson
fields, each one associated to up and down spin. We solve the
problem using as well the embedded cluster method (ECM), which is a powerfull tool to study strong correlated systems
\cite{anda}, and we compare the results of these two approaches. We study the
system for the two geometries mentioned above (side connected and
embedded). The value of the splitting results to be dependent on the
geometry of the circuit and has a strong dependence on the gate
potential applied to the QD and the coupling between the leads and
the QD. We also obtain the solution in the Hubbard I (HI)
aproximation \cite{March} that adequately describes the situation
for $T>T_k$.

\begin{figure}
\centering
\includegraphics[width=3.5in]{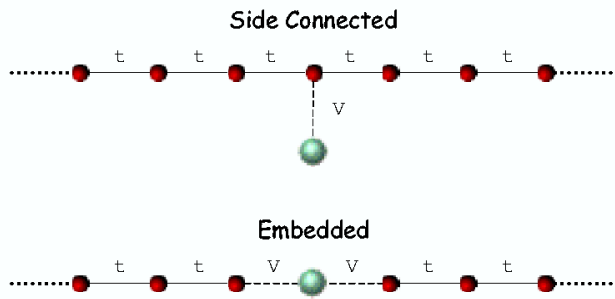}
\caption{Sketch of the geometry of the two topologies consider in this work.
Upper panel, SCG. Lower panel, EG.}

\label{fig:fig1}
\end{figure}

\noindent \noindent

The system is described by the Anderson impurity Hamiltonian
including the effect of an external magnetic field. We identify
three contribution to the total Hamiltonian: the leads ($H_L$), the
QD term ($H_C$) and the connections among the leads and the
dot($H_T$). $H_L$ is the Hamiltonian of two semi-infinite chains for
the EG configuration or of a one dimensional infinite chain for the
SCG case. The Hamiltonian $H_C$ is given by,

\begin{eqnarray}
{\hat H_C} & = &\sum_{\sigma} \epsilon_{\sigma} n_{0;\sigma} + U
n_{0;\uparrow} n_{0;\downarrow} \label{eq:H_C}
\end{eqnarray}

\noindent where the subindex $0$ denotes the QD site, $\epsilon_{\pm
\frac{1}{2}}= V_g \pm B$ is the spin dependent QD energy level,
$V_g$ is the gate potential at the QD site and $U$ the Coulomb
repulsion that, for simplicity, is supposed to be infinite. We have
adopted $|g\mu_B| = 1$. Finally,

\begin{eqnarray}
{\hat H_T} & = &\sum_{i;\sigma}\left[V a^{\dagger}_{i;\sigma}
a_{0\sigma}\right] + c.c.
\label{eq:H_T}
\end{eqnarray}

\noindent where $i$ is $-1$ and $1$ for EG or $i=1$ for the SCG. In
the first case $-1$ and $1$ denotes the first site of the left and
right semi-infinite one dimensional leads. In the second case $1$ is
the site of an infinite one dimensional lead to which the QD is
connected and $V$ is the hopping probability(see Fig. 1). 
All the energies are expressed in units of
the hopping among the leads sites $t$, $D$ denotes the bandwidth of
the leads and the Fermi energy is $E_f=0$.

In the new slave-boson formulation we define a boson field associate to each electron spin. The fermion operators
$a^{\dagger}_{0;\sigma}$ is written as the product, $a^{\dagger}_{0;\sigma}=c^{\dagger}_{0;\sigma} b^{\dagger}_{\sigma}$,
where $c$ and $b$ denote the quasi-fermion and boson
operators. Following a similar formulation to the standard one \cite{Hewson}, $H_T$ is transformed into

\begin{eqnarray}
{\hat H_T} & = &\sum_{i;\sigma}\left[V a^{\dagger}_{i;\sigma}
c_{0\sigma} b_{\sigma}\right] + c.c.
\label{eq:H_SB}
\end{eqnarray}

\noindent In order to avoid double occupancy at the QD, constraint
conditions have to be imposed. They are written,
\begin{eqnarray}
2 n_{0;\sigma} + b^{\dagger}_{-\sigma}b_{-\sigma}- b^{\dagger}_{\sigma}b_{\sigma}+
b^{\dagger}_{\sigma}b_{\sigma}b^{\dagger}_{-\sigma}b_{-\sigma}&=& 1.
\label{eq:Cons}
\end{eqnarray}

\noindent When both spins are equivalent, the above equations merge
into the constraint of the standard slave-boson formalism. The
Hamiltonian is treated in the mean-field approximation for the boson
fields (MFSB), substituting in equations (3) and (4) the operator
$b_{\sigma}$ by its mean-value $z_{\sigma} = <b_{\sigma}>$. The
mean-field versions of equations (4) are incorporated into the
Hamiltonian using the Lagrange multipliers $\gamma_{\sigma}$.
Minimizing the total energy with respect to these multipliers we
obtain two new equations,
\begin{eqnarray}
K V <a^{\dagger}_{1;\sigma}c_{0\sigma}> + z_{\sigma}^{2}(\gamma_{\sigma}+\gamma_{-\sigma})
+\gamma_{\sigma}-\gamma_{-\sigma}=0
\label{eq:Cons_2}
\end{eqnarray}
\noindent where $K=1$ for the SCG and $K=2$ for the EG. We solve
equations (5) self-consistently and obtain the Green function at the
QD and the transmission $T(E)$ across it ( $G = T(E_f)$).

We have chosen the parameters to obtain a similar Kondo resonance for the SCG and the EG configurations. They are
 $ V=0.35$ for the EG and $ V=0.65$ for the SCG. Increasing $V_g$ the
system goes from Kondo to mixed valence regime.

Fig.2 (upper panels) shows the transmission at zero field obtained
using the approximations MFSB and ECM. The effect of the Kondo
resonance is clear: $G=2$ or $G=0$ for the EG or the SCG
configurations respectively, and a resonance or anti-resonance width
roughly similar for both geometries and both methods of
calculations. When the magnetic field is increased the Kondo peak is
splitted and eventually the system goes out of resonance. In the
medium panels of Fig. 2 we show the transmission for $B=0.18$. The value of the
magnetic field splitting obtained by the two methods are very
similar, and for the ECM the results are almost independent of the
cluster size exactly diagonalized.

\begin{figure}
\centering
\includegraphics[width=3.5in,height=2in]{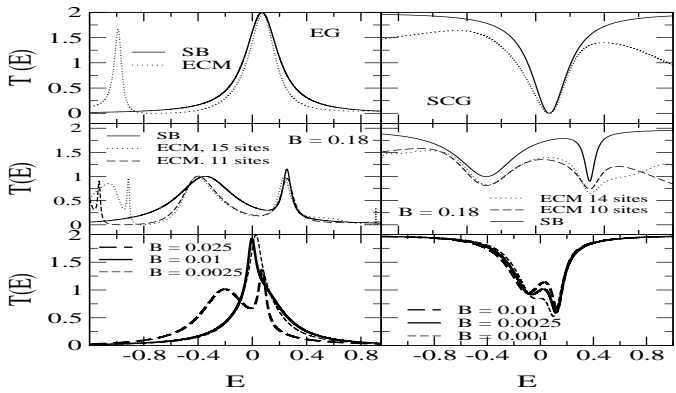}
\caption{Left column EG and right column SCG. Upper panels:
transmission as a function of energy for $B=0$. Continuous line SB,
broken line ECM(11 sites). Medium panels: the same as above but for
$B=0.18$ and two different cluster sizes. Lower panels: the same as
above for three different magnetic fields $B_1 < B_2 < B_3$ obtained
with MFSB. For more details see the text.}
\label{fig:fig4}
\end{figure}

To study how this splitting develops we write down the expression
for the Green function at the QD. For the MFSB solution we have,

\begin{eqnarray}
G_{0;\sigma}(E) =  \frac{1}{E-\epsilon_{\sigma}-2\gamma_{\sigma}+K(Vz_{\sigma})^{2}
g_L(E)} \label{eq:green}
\end{eqnarray}

\noindent Here $g_L(E)$ is the undressed Green function of the lead.
From equation (\ref{eq:green}) we can obtain $E_{\sigma}$, the values of the
renormalized energies of the QD level. They are,

\begin{eqnarray}
E_{\sigma} = \frac{\epsilon_{\sigma}+2\gamma_{\sigma}}{1-((K-1)Vz_{\sigma}/t)^{2}}.
\label{eq:energy}
\end{eqnarray}

\noindent The term $ \Gamma_{\sigma} =
\frac{2\gamma_{\sigma}}{1-(K-1)(Vz_{\sigma})^{2}}$ can be
interpreted as the spin dependent renormalization of the resonance
level by the interaction. Depending on the geometry ($K = 1$ or
$2$) there is also a multiplicative correction in the term
$\epsilon_{\sigma}$. The ZS is given by the difference $\Delta
=E_{\downarrow}-E_{\uparrow}$,

\begin{eqnarray}
\Delta = \frac{V_g+B+2\gamma_{\downarrow}}{1-(K-1)(V z_{\downarrow})^{2}}-\frac{V_g-B+
2\gamma_{\uparrow}}{1-(K-1)(V z_{\uparrow})^{2}}.
\label{eq:delta}
\end{eqnarray}

In Fig. 3 and 4 (lower panel) we show the parameters of the problem
as a function of the magnetic field. For both configurations
$z_{\uparrow} \rightarrow 1$ ,$z_{\downarrow}\rightarrow 0$ when
$B\rightarrow\infty$. $\Delta$ and
$\Delta_{\Gamma}=\Gamma_{\downarrow}-\Gamma_{\uparrow}$ show a rapid
increase at low fields, more evident for the SCG case.
$\Delta_{\Gamma}$ results to be almost constant for a large range of
values of the magnetic field. As a consequence, $\Delta(B)$ tends
asymptotically to a linear function. The slope of this linear
function is qualitatively the same as for the non interacting regime
(for the EG configuration there is a small correction proportional
to $V^2$ with respect to the non-interacting case). For low fields,
$\Delta$ rises due to the Zeeman effect but also because the local
level at the QD suffers a spin dependent renormalization as the
magnetic field progressively quenches the LSF, giving an additional
contribution to the splitting. The differential magnetic
susceptibility (DMS) is strongly enhanced for low magnetic fields, 
(see the curve $n_{0,\uparrow}-n_{0,\downarrow}= \Delta_n$ in figures
3,4). For larger fields, when the spin fluctuations are quasi
totally suppressed, $\Delta$ recovers its linear function appearance,
but shifted up with respect to the origin, due to the Coulomb
interaction acting on each spin level. Lowering $V_g$, the
Y-intercept coordinate of this linear function is increased,
although the slope does not change. For a fixed value of the field
$B$, $\Delta$ results to depend linearly on $V_g$ with a slope
$\alpha$ almost independent of $B$ (see the inset of figures 3 and
4). This implies that, as known \cite{amge05}, $\Delta$ increases
logaritmically by lowering the Kondo temperature since $V_g \sim
(V^{2}/D) log(D/T_k)$. As $\Delta>\Delta_B$, the slope $\alpha$ is
larger than the value predicted in \cite{co00}, and in agreement
with recent experiments \cite{koam04,amge05}.
Increasing $V_g$, we enter into the mixed valence regime and
$\Delta$ recovers a  nearly linear evolution that almost extrapolates
to the origin. In this regime $\Delta_{\Gamma}$ is not constant anymore
but increases monotonically with the field. As a consequence, the
slope of the linear relationship between $\Delta$ and $B$ is
renormalized  for both geometries with respect to the
non-interacting case. In spite of this, it is important to
realize that $\Delta$ is always lower than the value it acquires in the
Kondo regime.

\begin{figure}
\centering
\includegraphics[width=2.5in,height = 3in]{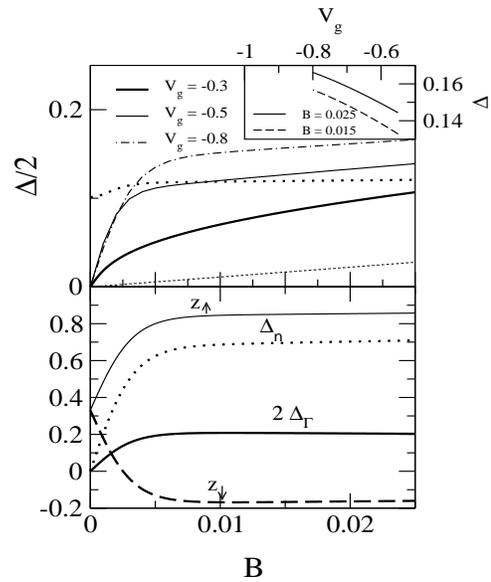}
\caption{Results for EG. Upper panel: The function $\Delta(B)$ for
three different values of $V_g$.  The semi-sum of the resonance
widths $W$, for $Vg=-0.5$, is shown with thick point line. The
equation $W(B)=\Delta(B)$ defines $B_c$. The point line shows
$\Delta$ using the Hubbard I approximation. Lower panel: evolution
of the parameters with $B$, for $V_g = -0.8$. The inset shows
$\Delta(V_g)$ for two magnetic fields.} \label{fig:fig4}
\end{figure}

To determine when the ZS can be experimentally detected the relevant
quantity to be study is $W$, the semi-sum of the resonance width of
each Kondo peak. It can be expressed as,
\begin{eqnarray}
W= K \left[\frac{(Vz_{\downarrow})^{2}}{1-(K-1)(V z_{\downarrow})^{2}}+\frac{(Vz_{\uparrow})^{2}}{1-(K-1)
(V z_{\uparrow})^{2}}\right].
\label{eq:width}
\end{eqnarray}
The ZS is not detectable if $\Delta < W$. The value of the critical
magnetic field $B_c$ satisfies the equation $\Delta(B_c)=W(B_c)$. In
figure 3 and 4 we plot $W(B)$ for $V_g = -0.5$ and $V_g= -0.4$,
respectively. The values of $W$ obtained from (9) at $B=0$ are the
same to those obtained with the usual formula $T_k $=$D
exp(-xV_g/2KV^2)$\cite{Hewson} where $x$ is $\pi$ for EG or $2\pi$
for SGC. The $B_c$ obtained for each configuration corresponds to a
$\Delta(B_c)<< T_k$ in line with the experimental observation. Note that $B_c$ is appreciably smaller for the SCG than
for the EG topology because in this case $\Delta$ has
a negative contribution comming from $V_g$ (8), which is absent for the SCG, and also $\Delta_{\Gamma}$ has a more rapid 
increasing with $B$ for the SCG. In figure 2 (lower pannels) it is shown how the splitting develops for each geometry.

We have also solved the problem in the Hubbard I approximation
\cite{March}. It describes the situation for $T>T_K$, ${\it i.e}$
when the system is outside the Kondo regime. The local Green
function results to be:
\begin{eqnarray}
G_{0;\sigma}(E) =  \frac{1}{E-\epsilon_{\sigma}+K V^{2}(1-n_{-\sigma})g_L(E)}
\label{eq:greenHI}
\end{eqnarray}

\begin{figure}
\centering
\includegraphics[width=2.5in,height = 3in]{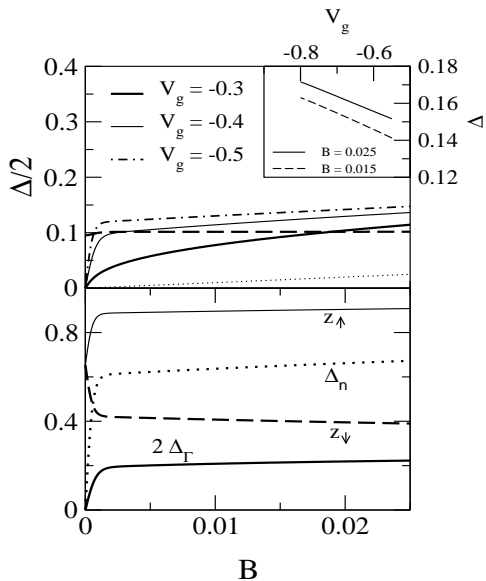}
\caption{Results for the SCG. Upper panel: The function $\Delta(B)$
for three different values of $V_g$. The semi-sum of the resonance
widths $W$, for $Vg=-0.5$, is shown with thick trace line. The
equation $W(B)=\Delta(B)$ defines $B_c$. The point line shows
$\Delta$ using the Hubbard I approximation. Lower panel: evolution
of the parameters with $B$, for $V_g = -0.5$. The inset shows
$\Delta (V_g)$ for two magnetic fields.} \label{fig:fig3}
\end{figure}

We can determine the position of the resonances as a function of the
external field \cite{March}. $\Delta$ results to be:
\begin{eqnarray}
\Delta = \frac{V_g+B}{1-(K-1)V^{2}(1-n_{\uparrow})}-\frac{V_g-B}{1-(K-1)V^{2}(1-n_{\downarrow})}.
\label{eq:deltaHI}
\end{eqnarray}
The function $\Delta(B)$ is plotted in figures 3 and 4. The
remarkable fact is that for the SCG configuration the slope is equal
to that of the low temperature solutions obtained with MFSB, while
for the EG there is a small renormalization of it. This shows that
the essential difference between the low and  high temperature
regime as far as the ZS is concerned is the existence of a non-zero
Y-intercept coordinate at low temperatures as has been seen in the
experiment. In spite of the similar behavior of the slope of the
function $\Delta(B)$, the magnetization in this high temperature
regime is very different. The DMS at zero field is much less at high temperature than
at low temperature.

Summarizing, we have studied the magnetic response of a QD connected
with two different geometries SCG and EG. To do so we developed a
slave-boson formalism that takes into account a different boson
field for each spin. In the Kondo regime there is a critical
magnetic field $B_c$ above which the Kondo resonance can be seen
splitted. $B_c$ depends upon the topology of the system, being
larger for the EG than for the SCG configuration. The Zeeman energy
associated to this field is lower that the Kondo temperature. For
$B<B_c$, $\Delta$ suffers a non-linear and rapid growth with the
magnetic field that occurs due to the spin dependent local energy
Coulomb renormalization and the quenching of the LSF induced by the
magnetic field. When the spin fluctuations are completely quenched,
$B>B_c$, the spin splitting is proportional to the magnetic field
with a slope independent on temperature and gate potential. The
Y-interception coordinate is non zero and increases when the gate
potential is reduced. At a fixed magnetic field, $\Delta(V_g)$
results to be a linear function with a $B$ independent slope. The
value of $\Delta$ is always greater than $ \Delta_B$ for $T<T_k$
due to the spins correlations implied in the Kondo effect. Our
results are in good agreement with recent experimental results.

\acknowledgments

\noindent Financial support by the argentinian CONICET and UBA (grant UBACYT x210) and the spanish programn
Ramon y Cajal of MCyT and Brazilian agencies, FAPERJ, CNPq, and CNPq(CIAM project) are gratefully acknowledged.
Also funding from Generalitat Valenciana, (grant GV05/152), and by the spanish MCYT (grants MAT2005-07369-C03-01 and 
NAN2004-09183-C10-08) are acknowledged.

\end{document}